# BAYESIAN MODELS TO ADJUST FOR RESPONSE BIAS IN SURVEY DATA FOR ESTIMATING RAPE AND DOMESTIC VIOLENCE RATES FROM THE NCVS[1]

BY QINGZHAO YU, ELIZABETH A. STASNY AND BIN LI

*Louisiana State University and Ohio State University*

It is difficult to accurately estimate the rates of rape and domestic violence due to the sensitive nature of these crimes. There is evidence that bias in estimating the crime rates from survey data may arise because some women respondents are "gagged" in reporting some types of crimes by the use of a telephone rather than a personal interview, and by the presence of a spouse during the interview. On the other hand, as data on these crimes are collected every year, it would be more efficient in data analysis if we could identify and make use of information from previous data. In this paper we propose a model to adjust the estimates of the rates of rape and domestic violence to account for the response bias due to the "gag" factors. To estimate parameters in the model, we identify the information that is not sensitive to time and incorporate this into prior distributions. The strength of Bayesian estimators is their ability to combine information from long observational records in a sensible way. Within a Bayesian framework, we develop an Expectation-Maximization-Bayesian (EMB) algorithm for computation in analyzing contingency table and we apply the jackknife to estimate the accuracy of the estimates. Our approach is illustrated using the yearly crime data from the National Crime Victimization Survey. The illustration shows that compared with the classical method, our model leads to more efficient estimation but does not require more complicated computation.

**1. Introduction.** Rape and domestic violence rates are difficult to estimate because of difficulties in collecting data on these crimes. Annual rape incidence rates in the U.S. obtained from police statistics, reported through

Received March 2007; revised January 2008.

[1]Supported in part by award number 93-IJ-CX-0050 from the National Institute of Justice, Office of Justice Programs, U.S. Department of Justice, and by the Center for Survey Research Summer Fellowship, OSU.

*Key words and phrases.* Categorical data analysis, contingency table, EMB algorithm, incompletely classified data, Jackknife method, panel survey, survey mode.

the Uniform Crime Report (UCR), were estimated to be 0.3 per 1,000 persons in females age 12 and older [Bureau of Justice Statistics (2002)]. But the majority of both rape and domestic violence incidents are not reported to police. Data from the National Women's Study, a longitudinal telephone survey of a national household probability sample of women at least 18 years of age, show that 683,000 women were forcibly raped each year and that 84% of rape victims did not report the offense to the police (CDC's National Center of Injury Prevention and Control web site). Thus, we believe that the UCR underestimates the rates of such crimes. The National Crime Victimization Survey (NCVS) is the best national source for estimates of rates of rape and domestic violence in the United States. It includes crimes reported and not reported to the police. But NCVS rates might still be biased because: (a) the definition of "criminal" rape or domestic violence was left to the respondent who may define rape differently relative to the legal definition; (b) personal NCVS interviews may not be conducted confidentially (with the respondent alone); and (c) telephone interviews may not be sufficiently private. Leggett et al. (2003) reported that people are more likely to report in a face-to-face interview, compared with a telephone interview. In this paper we deal with respondent bias caused by (a) privacy concerns in telephone interviews and (b) the in-person interviews not conducted confidentially.

It is important to have accurate and reliable estimates of rape and domestic violence rates. The criminal justice community, for example, needs accurate estimates of these rates to evaluate how well it meets the needs of victims, and whether criminal justice interventions help reduce the rates of rape or domestic violence. Since the NCVS is a population-based data source for estimating crime rates, interventions to reduce rape (or domestic violence) can use the NCVS data to help evaluate their efficacy.

The NCVS survey can be conducted by phone or in-person. Other individuals are allowed to be present during the interview. Our research on the NCVS data (see Section 2.2) suggests that rape and domestic violence incidents are under-reported by women in telephone interviews, or if a spouse is present during the interview. We refer to these as the gag factors in reporting rape and domestic violence. Our research also indicates that the effect of gag factors in underreporting crimes is relatively constant over time. In the following we develop a Bayesian model that allows the use of both current data and previous information to estimate rates of rape and domestic violence while taking into account potential under-reporting of these crimes due to the gag effect. This paper presents a Bayesian model for estimating classification probabilities and probabilities of respondents' bias. Our models take into account the respondent bias caused by known influential factors. Moreover, in estimating the distributions of interest, we identify from the previous data the information that is not sensitive to time and use it to build prior distributions for some of the parameters, so that the previous information is efficiently used in estimation. We also provide computation methods that allow complicated posterior distributions to be explored through simple iterations. We use the jackknife in the Bayesian environment to calculate the accuracy of the estimates, and then compare the results to estimates obtained from the classical method. We illustrate our methodology by estimating rates of rape and domestic violence using the public NCVS data.

The background literature for our response-bias-adjusted Bayesian models is based on three lines of research. First, extensive research has been done on methods to capture information on respondent bias. Our model has its roots in the work of Stasny and Coker (1997), who identified the problem of response bias in self-reported crime and developed a model to adjust for the bias. In this paper we modify the model and incorporate historical information to improve estimation. The second line of background research concerns the extension of the EM algorithm to find empirical Bayesian estimates in analyzing contingency tables. See Little and Rubin (2002) for an introduction to EM algorithm; Carlin and Louis (2000) and Bishop, Fienberg and Holland (1975) for an introduction of empirical-Bayesian estimators of cell probabilities. We extend the EM algorithm to find empirical Bayesian estimates by adding a third step, the B step, in which we calculate empirical-Bayesian estimates based on current maximum likelihood estimates. We call this an Expectation-Maximization-Bayesian (EMB) algorithm. To our knowledge, this is the first proposal for such an algorithm, although there are Bayesian versions of EM by Gelman et al. (2004). EMB is useful in estimating survey classification and respondent bias caused by any influential factors when prior information is available. Third, we use the jackknife method [Efron (1987)] in the Bayesian environment to measure the accuracy of the estimates.

In the next section we provide a brief description of the NCVS and an exploratory analysis of the data. Section 3 presents a model that adjusts data for gag factors. Section 4 discusses efficiency gains using the Bayesian model and describes building prior distributions, the EMB algorithm and how to use the jackknife method in the Bayesian environment. Section 5 presents the results of analysis on NCVS. Finally, Section 6 points out some directions for future research.

## 2. The National Crime Victimization Survey and data.

2.1. *Survey design.* The National Crime Victimization Survey is administered by the US Census Bureau on behalf of the Bureau of Justice Statistics. The survey has been collecting data on personal and household victimizations since July 1972. It was formerly known as the National Crime Survey before its redesign in 1989 when the current survey methodology began systematic field testing. The first annual results from the redesigned survey were published in 1993 [Bureau of Justice Statistics (1995)]. There were some further changes to the survey after that, but the data collection procedure and instrument that provides important information to our analysis were consistent over these years. We use data from 1993 to the most currently available data online. The NCVS is the primary source of national-level information on victimizations, including not only crimes reported to the police, but also those not reported to law enforcement authorities. We briefly describe the NCVS in the rest of this subsection. Additional information on the design and history of the NCVS is provided, for example, by the U.S. Department of Justice and Bureau of Justice Statistics (2001) and Stasny and Coker (1997).

The NCVS is comprised of a stratified, multi-stage, cluster sample of housing units (HUs). This ongoing survey seeks to obtain a representative sample of individuals 12 years of age and older living in households or group quarters within the United States. Semi-annual data on the frequency, characteristics and consequences of criminal victimization are collected from approximately 49,000 households comprising about 100,000 persons. The NCVS uses a rotating panel design whereby a sampled HU is maintained in the sample for seven panels with interviews conducted at six-month intervals. The first interview conducted within a household is considered a bounding interview, which is not published but is used as a control to avoid duplicate reporting of an incident. New households rotate into the sample on an ongoing basis. During the interview, individuals are asked about crimes committed against them or against the household (HH) in the past six months. The crimes are categorized as personal (which includes rape/sexual assault, robbery, aggravated/simple assault and personal larceny) or property (which includes burglary, auto or motor vehicle theft, theft and vandalism) related. Crimes not covered include kidnapping, murder, shoplifting and crimes that occur at places of business. The survey instrument is composed of a screening section and an incident report. A single HH respondent is asked a series of six screening questions to elicit information on crimes committed against the HH (e.g., burglary, larceny, motor vehicle theft). Next, an eleven-question screener is used to elicit information from each individual in the HH concerning personal crimes committed against that individual. If any screening question elicits a positive response, an incident report is filled out. The report is designed to obtain detailed data on the characteristics and circumstances of the crime, such as the month, time, location of the incidence, relationships between victim and offender, offender characteristics, self-protective actions, type of property lost, whether crime was reported to the police, consequences of the victimization and offender use of weapons, drugs or alcohol.

The initial NCVS interview at a housing unit must be conducted in person. The subsequent survey contacts at the same address could be conducted either through telephone or face-to-face. Primarily for cost reasons, phone contacts are emphasized in the later interviews; a face-to-face interview is conducted only when it is inefficient or infeasible to make contact by phone. Since respondents are asked to describe the victimization, the lack of privacy can influence responses during a telephone interview. Ideally, a personal interview is conducted and the interviewer and respondent are alone during the interview. In our data, approximately 45% of personal interviews are conducted alone with the respondent. However, this is not always possible. In neither the phone nor personal surveys are interviewers instructed to establish a private interview setting. During the face-to-face interview, if the respondent is not alone, the interviewer indicates on the questionnaire who else is present. When a telephone interview is conducted, the respondent is not asked so we have no information about whether other individuals are present.

Although there are limitations in using the NCVS to estimate rates of both rape and domestic violence, this data set is the only on-going large and nationally representative survey to ask individuals directly whether they have been victims of specific crimes.

2.2. *The survey data.* Our analysis data set includes all women 16 years of age or older in the NCVS data base, with the exception of proxy interviews for the years from 1993 to 2004. We use the NCVS data from 1993 to 1997 as prior information and conduct the analysis on data from 1998 to 2004. It is necessary to combine information from a number of years because rape and domestic violence are relatively rare events. We combine the data for the years 1993 to 1997 and for the years 1998 to 2004 since a descriptive analysis shows that although the crime rates increases from the first period to the second period, the crime rates in each period are almost constant. We recognize that there exists a potential correlation in responses from the same woman over time. But this correlation should not present a significant problem in our analysis since: (a) the survey is controlled so that no crime is repeatedly recorded; (b) the data are collected so that people have the same chance of being included in the sample; and (c) an analysis is also performed using weighted data, which should better represent the population of interest.

The raw, unweighted data from 1998 to 2004 by type of crime, type of interview and who was present during personal interviews are presented in Table 1. We divide the crimes into four groups: rape, domestic violence, other assault and personal larceny. Rape, attempted rape and sexual assault are categorized as rape. The types of crimes included in the domestic violence and other assault categories are exactly the same. If the offender is an intimate, then the assault is categorized as domestic violence. Otherwise, it is categorized as other assault. An intimate is defined as a spouse,

6    Q. YU, E. A. STASNY AND B. LI

ex-spouse, boyfriend or ex-boyfriend. Personal larceny includes purse snatching and pocket picking. We exclude verbal crimes in our analysis because these kinds of crimes are hard to define. Table 1 lists frequencies and rates of crimes reported by interviewers.

Four categories are used to describe who was present during the personal interview: (1) a spouse and no one else (labeled Spouse), (2) a spouse and at least one other person (Spouse and Other), (3) at least one person but no spouse (Other), and (4) no one else present (Alone). As mentioned previously, we do not know who is present with the responding woman during telephone interviews. Note that the raw data report crime rates per 1,000 women interviewed as follows: 0.79 rapes, 1.66 incidence of domestic violence, 4.14 other assaults and 0.60 incidence of personal larceny.

If we consider the rates of various crimes in Table 1 by type of interview, we find that there are some large differences as shown in Figure 1. Except for personal larceny, more crimes were reported in personal interviews than in telephone interviews. For example, rape was reported at a rate 1.45 times higher in personal interviews compared to telephone interviews; domestic violence was reported at a rate 1.33 times higher and other assault was reported at a rate 1.11 times higher. Thus, the telephone interview appears to have a gag effect in the reporting of crimes.

From Table 1, we also note that rates of reported crimes in personal interviews depend on who was present during the interview. For personal interviews, we compare other situations to the case in which the woman was interviewed alone, since that is considered the ideal case. Compared with a woman who was interviewed alone, rape was reported about one-fifth as frequently when a spouse was present (either with or without others). Considering domestic violence, the incident was reported approximately one-tenth as frequently if a spouse was present. The other assault category was reported 1/2.93 as frequently, and personal larceny was reported 1/1.84 as frequently if a spouse was present during the interview. Part of this reporting differential may be explained by the protection offered by having a spouse in the household. All rates of reported crimes were lower when a spouse was present, however, the reported rates for rapes and domestic violence are significantly lower than those obtained when the woman was interviewed alone. We therefore believe that the spouse being present during an interview has a differential influence on the reporting of rape and domestic violence. Rape is under-reported because it is a sensitive crime, while domestic violence is under-reported not only because it is a sensitive crime but also because the offender may be present.

TABLE 1
*Frequencies and rates of crimes reported by settings of the interviews: NCVS 1998–2004*

| | Interviews | | Numbers of incidents reported by type of personal crimes (rates per 1000 interviews) | | | | |
|---|---|---|---|---|---|---|---|
| Type of Interview | Who present | Number | Rape | Domestic violence | Other assault | Personal larceny | No crime reported |
| Telephone | Unknown | 412339 | 288 | 516 | 1539 | 270 | 409726 |
| | | | (0.70) | (1.25) | (3.73) | (0.65) | (993.66) |
| | Spouse | 24063 | 4 | 12 | 45 | 10 | 23992 |
| | | | (0.17) | (0.50) | (1.87) | (0.42) | (997.05) |
| | Spouse | 14322 | 5 | 2 | 29 | 2 | 14284 |
| Personal | and Other | | (0.35) | (0.14) | (2.02) | (0.14) | (997.35) |
| | Other | 48916 | 6 | 162 | 350 | 18 | 48320 |
| | | | (1.35) | (3.31) | (7.16) | (0.37) | (987.82) |
| | Alone | 71708 | 86 | 256 | 403 | 41 | 70922 |
| | | | (1.20) | (3.57) | (5.62) | (0.57) | (989.04) |
| All personal | | 159009 | 161 | 432 | 827 | 71 | 157518 |
| | | | (1.01) | (2.72) | (5.20) | (0.45) | (990.62) |
| All interviews | | 571348 | 449 | 948 | 2366 | 341 | 567244 |
| | | | (0.79) | (1.66) | (4.14) | (0.60) | (992.82) |

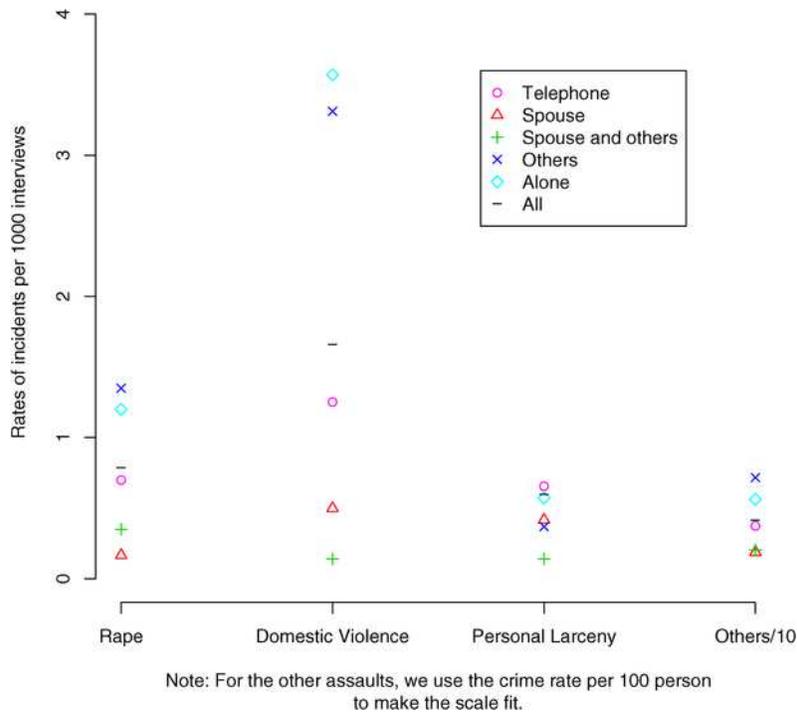

FIG. 1. *Comparison of crime rates reported by women by type of interview and who was present during interview.*

TABLE 2
*Observed data (frequencies) for reporting rape, domestic violence and other crimes from 1998 to 2004*

|  | Personal interview | | Telephone interview |
|---|---|---|---|
|  | **Spouse present** | **Spouse not present** |  |
| Rape | 9 | 152 | 288 |
| Domestic violence | 14 | 418 | 516 |
| Other assault | 74 | 753 | 1539 |
| Personal larceny | 12 | 59 | 270 |
| No crime | 38276 | 119242 | 409726 |

This exploratory analysis of the raw data suggests that there is response bias in the NCVS related to type of interview (personal versus telephone) and who was present during the interview. In the later case, the bias is particularly large in the reporting of the sensitive crimes of rape and domestic violence. Thus, we consider a model that allows us to adjust the estimates for those crime rates according to the possible response bias.

**3. A model for response bias adjustment.** In our model we classify the data according to the type of crime, presence of the spouse (either with or without others) during the interview and whether the interview was conducted by telephone or face-to-face. The crimes are classified into five categories: (1) rape and possibly some other crime, (2) domestic violence, not rape but possibly some other crime, (3) other assault except for rape and domestic violence, (4) personal larceny except for all kinds of assault, and (5) no personal crime reported. The unweighted and weighted data are summarized in Table 2 and Table 3 respectively. Weighted data are used in the analysis to account for the sample design and socio-economic indicators.

Tables 2 and 3 do not present the "truth." Some women prefer not reporting crimes under certain circumstances. We want to build a model taking into account the circumstances of the interview to obtain more accurate estimates of the rates of rape and domestic violence. We suspect that our estimates still underestimate the true rates since some women would never report some incidents under any circumstances. We use the information available on factors believed to provide gag effects to improve estimation.

TABLE 3
*Weighted-adjusted data (frequencies) for reporting rape, domestic violence and other crimes from 1998 to 2004*

|  | Personal interview | | Telephone interview |
|---|---|---|---|
|  | **Spouse present** | **Spouse not present** |  |
| Rape | 8.93 | 177.65 | 338.66 |
| Domestic violence | 13.19 | 492.70 | 565.67 |
| Other assault | 77.08 | 867.77 | 1673.10 |
| Personal larceny | 10.28 | 57.99 | 263.93 |
| No crime | 37234.91 | 121675.28 | 407890.86 |

In our model we assume a woman may be gagged from reporting a crime for two reasons: a spouse was present, or the interview was conducted over the telephone. We further assume that the spouse's presence only influences the reporting of rape and domestic violence, whereas conducting the interview over the phone may influence all crimes except for personal larceny. To ensure the model is identifiable, we impose a hierarchy on the reasons for not reporting crimes. Namely, we assume that the presence of a spouse dominates the use of a telephone interview in determining whether or not a woman reports such an incident. So if we could only observe it, the complete data underlying Table 2 would tell us if a crime actually occurred and whether it was reported. If a crime was not reported, the unobserved complete data could tell us if the crime was not reported because of the presence of a spouse, or was not reported because the interview was conducted over the telephone. The form of the complete (but unobserved) data for reporting crimes is shown in Table 4.

Note that some outcomes are impossible, for example, not reporting because the spouse is present when the woman was interviewed without the spouse's presence. Such impossible outcomes are denoted by a dash in Table 4.

We now present a model to analyze the probabilistic relationship between the underlying complete data and the observed data. The following notation is employed:

$\pi =$ probability of a telephone interview,

$1 - \tau =$ probability of crimes not reported because of telephone interview,

$1 - \rho =$ probability of rape not reported because spouse is present,

$1 - \delta =$ probability of domestic violence not reported because spouse is present,

$\omega_{ij} =$ probability of crime status $i$ and interview status $j$,

where $j = 1$ if spouse is present, 2 if spouse is not present; $i = 1$ if rape, 2 if domestic violence, 3 if other assault, 4 if personal larceny, 5 if no crime.

We fit a model that assumes the independence between crimes and spouse presence, so that $\omega_{ij} = c_i \cdot s_j$, where $c_i$ denotes the probability of each type of crime and $s_j$ denotes the probability of the spouse present during the interview. Under the assumptions described above, the probabilities underlying the unobserved complete data are shown in Table 5.

Table 4
*Form of unobserved complete data*

| **Personal interview** | | | |
|---|---|---|---|
| | | **Spouse** | |
| | | **Present** | **Not present** |
| Rape | Reported | $y_{1111}$ | $y_{1112}$ |
| | Not reported—spouse present | $y_{1121}$ | — |
| Domestic | Reported | $y_{1211}$ | $y_{1212}$ |
| violence | Not reported—spouse present | $y_{1221}$ | — |
| Other assault | Reported | $y_{1311}$ | $y_{1312}$ |
| Personal larceny | Reported | $y_{1411}$ | $y_{1412}$ |
| No crime | Reported | $y_{1511}$ | $y_{1512}$ |

| **Telephone interview** | | | |
|---|---|---|---|
| | | **Spouse** | |
| | | **Present** | **Not present** |
| Rape | Reported | $y_{2111}$ | $y_{2112}$ |
| | Not reported—spouse present | $y_{2121}$ | — |
| | Not reported—phone interview | $y_{2131}$ | $y_{2132}$ |
| Domestic | Reported | $y_{2211}$ | $y_{2212}$ |
| violence | Not reported—spouse present | $y_{2221}$ | — |
| | Not reported—phone interview | $y_{2231}$ | $y_{2232}$ |
| Other | Reported | $y_{2311}$ | $y_{2312}$ |
| assault | Not reported—phone interview | $y_{2331}$ | $y_{2332}$ |
| Personal larceny | Reported | $y_{2411}$ | $y_{2412}$ |
| No crime | Reported | $y_{2511}$ | $y_{2512}$ |

In the observed data, some of the cells from the complete data are collapsed [see, e.g., Chen and Fienberg (1974, 1976)]. Hence, we observe only sums of several cells rather than all 30 possible cells represented in the complete-data table. Table 6 presents the notation for the observed data table and indicates which cell counts from the complete data table are summed together to create the observed data. The probabilities underlying the observed data are similarly just the sums of the probabilities underlying the unobserved complete data and are shown in Table 7.

## 4. Bayesian inference for the bias-adjusting model.

4.1. *Why use a Bayesian model.* Our goal is to estimate the $\pi, \tau, \rho, \delta$ and $\omega$ based on observations y in the model described in Section 3. We adopt a Bayesian perspective that treats the parameters $\tau, \rho$ and $\delta$ as random variables to account for the uncertainties in $\tau, \rho$ and $\delta$. We incorporate prior information provided by the NCVS data from 1993 to 1997. Notice that we do not estimate $\pi$ in this way since $\pi$, the proportion of telephone interviews, is a fixed value and cannot be influenced by the prior surveys. We would not use the prior information to estimate $\omega$ since the crime rates change considerably between the 93–97 and the 98–04 time periods. To demonstrate the difference, we build two models: Model A that assumes equal crime rates between time periods, and Model B that does not make this assumption. The likelihood-ratio model comparison test shows an improvement of 740 in the $G^2$ with 4 degrees of freedom of model B over model A.

TABLE 5
*Probabilities underlying unobserved complete data*

| | **Personal interview** | | |
|---|---|---|---|
| | | **Spouse** | |
| | | **Present** | **Not present** |
| Rape | Reported | $(1-\pi)\rho\omega_{11}$ | $(1-\pi)\omega_{21}$ |
| | Not reported—spouse present | $(1-\pi)(1-\rho)\omega_{11}$ | — |
| Domestic | Reported | $(1-\pi)\delta\omega_{12}$ | $(1-\pi)\omega_{22}$ |
| violence | Not reported—spouse present | $(1-\pi)(1-\delta)\omega_{12}$ | — |
| Other assault | Reported | $(1-\pi)\omega_{13}$ | $(1-\pi)\omega_{23}$ |
| Personal larceny | Reported | $(1-\pi)\omega_{14}$ | $(1-\pi)\omega_{24}$ |
| No crime | Reported | $(1-\pi)\omega_{15}$ | $(1-\pi)\omega_{25}$ |
| | **Telephone interview** | | |
| | | **Spouse** | |
| | | **Present** | **Not present** |
| Rape | Reported | $\pi\tau\rho\omega_{11}$ | $\pi\tau\omega_{21}$ |
| | Not reported—spouse present | $\pi(1-\rho)\omega_{11}$ | — |
| | Not reported—phone interview | $\pi\rho(1-\tau)\omega_{11}$ | $\pi(1-\tau)\omega_{21}$ |
| Domestic | Reported | $\pi\tau\delta\omega_{12}$ | $\pi\tau\omega_{22}$ |
| violence | Not reported—spouse present | $\pi(1-\delta)\omega_{12}$ | — |
| | Not reported—phone interview | $\pi\delta(1-\tau)\omega_{12}$ | $\pi(1-\tau)\omega_{22}$ |
| Other | Reported | $\pi\tau\omega_{13}$ | $\pi\tau\omega_{23}$ |
| assault | Not reported—phone interview | $\pi(1-\tau)\omega_{13}$ | $\pi(1-\tau)\omega_{23}$ |
| Personal larceny | Reported | $\pi\omega_{14}$ | $\pi\omega_{24}$ |
| Personal larceny | Reported | $\pi\omega_{15}$ | $\pi\omega_{25}$ |

The influence of the "gag" factors on crime reports might also change over the years, but this change is not significant as is shown in the following tests. This might result from the facts that features of the survey instrument such as the question order, wording of the questions and the gateway questions are consistent over the years. Table 8 shows the number of rapes reported by women when a spouse is or is not present during the interview in the periods 1993 to 1997 and 1998 to 2004. We use the Breslow–Day method to test the homogeneity of the odds ratio in the contingency tables and get a test statistic $\chi^2 = 0.2649$ with 1 degree of freedom, and a p-value of 0.6068, which means that the influence of spouse-presence on rape reporting is not significantly different for the two periods. We do the same test on the reporting of domestic violence and draw the same conclusion (p-value = 0.3049). To test the gag effect of telephone interview on crime reports, we cannot use the same method, because the effect from "who was present during the interview" is confounded in the data. We use the data from years 1993 to 1997 efficiently through the use of empirical-Bayesian estimates. The method is described in the next section and in the results (Section 5), we clearly show that our method greatly improves estimation without increasing computational load.

TABLE 6
*Form of observed data*

| | **Personal interview** | |
|---|---|---|
| | **Spouse present** | **Not present** |
| Rape reported | $x_{111} = y_{1111}$ | $x_{112} = y_{1112}$ |
| Domestic violence reported | $x_{121} = y_{1211}$ | $x_{122} = y_{1212}$ |
| Other assault reported | $x_{131} = y_{1311}$ | $x_{132} = y_{1312}$ |
| Personal larceny reported | $x_{141} = y_{1411}$ | $x_{142} = y_{1412}$ |
| No crime reported | $x_{151} = y_{1121} + y_{1221} + y_{1511}$ | $x_{152} = y_{1512}$ |
| | **Telephone interview** | |
| Rape reported | $x_{21} = y_{2111} + y_{2112}$ | |
| Domestic violence reported | $x_{22} = y_{2211} + y_{2212}$ | |
| Other assault reported | $x_{23} = y_{2311} + y_{2312}$ | |
| Personal larceny reported | $x_{24} = y_{2411} + y_{2412}$ | |
| No crime reported | $x_{25} = y_{2121} + y_{2131} + y_{2132} + y_{2221} + y_{2231} +$ | |
| | $\quad y_{2232} + y_{2331} + y_{2332} + y_{2511} + y_{2512}$ | |

TABLE 7
*Probabilities underlying the observed data*

| | **Personal interview** | |
|---|---|---|
| | **Spouse** | |
| | **Present** | **Not present** |
| Rape reported | $(1-\pi)\rho\omega_{11}$ | $(1-\pi)\omega_{21}$ |
| Domestic violence reported | $(1-\pi)\delta\omega_{12}$ | $(1-\pi)\omega_{22}$ |
| Other assault reported | $(1-\pi)\omega_{13}$ | $(1-\pi)\omega_{23}$ |
| Personal larceny reported | $(1-\pi)\omega_{14}$ | $(1-\pi)\omega_{24}$ |
| No crime reported | $(1-\pi)(1-\rho)\omega_{11}+$ | $(1-\pi)\omega_{25}$ |
| | $(1-\pi)(1-\delta)\omega_{12} + (1-\pi)\omega_{15}$ | |
| | **Telephone interview** | |
| Rape reported | $\pi\tau\rho\omega_{11} + \pi\tau\omega_{21}$ | |
| Domestic violence reported | $\pi\tau\delta\omega_{12} + \pi\tau\omega_{22}$ | |
| Other assault reported | $\pi\tau\omega_{13} + \pi\tau\omega_{23}$ | |
| Personal larceny reported | $\pi\tau\omega_{14} + \pi\tau\omega_{24}$ | |
| No crime reported | $\pi(1-\rho)\omega_{11} + \pi\rho(1-\tau)\omega_{11} + \pi(1-\tau)\omega_{21}+$ | |
| | $\pi(1-\delta)\omega_{12} + \pi\delta(1-\tau)\omega_{12} + \pi(1-\tau)\omega_{22}+$ | |
| | $\pi(1-\tau)\omega_{13} + \pi(1-\tau)\omega_{23} + \pi\omega_{14} + \pi\omega_{24}$ | |

4.2. *Incorporating prior information.* Since the parameters we want to estimate are probabilities, we choose the Dirichlet distribution as the prior distribution since it is the natural conjugate family of prior distributions for the multinomial distribution [Berger (1985)]. We now use the data from 1993 to 1997 to form prior distributions. Let $X = (x_1, x_2, \ldots, x_t)$ be multinomial distributed with parameters $n$ and $P = (p_1, p_2, \ldots, p_t)$. Let the prior distribution for $P$ be Dirichlet with parameters $(\beta_1, \beta_2, \ldots, \beta_t)$. Then the posterior distribution is also Dirichlet with parameters $B + X = (\beta_1 + x_1, \beta_2 + x_2, \ldots, \beta_t + x_t)$. Under the square error loss function, the Bayesian estimator of $P$ is the posterior mean. If we set $k = \sum_{i=1}^{t} \beta_i$ and $\lambda_i = \beta_i/k, \Lambda = (\lambda_1, \lambda_2, \ldots, \lambda_t)$, which is a one-to-one transformation from the parameters $(\beta_1, \beta_2, \ldots, \beta_t)$, then the estimate for $P$ is

$$E(P|k, \Lambda, X) = \frac{n}{n+k} \times \frac{X}{n} + \frac{k}{n+k} \times \Lambda.$$

We want to find proper parameters for the Dirichlet distribution to incorporate the information from the 1993 to 1997 data. We know that the prior means of the $p_i$ are given by $E(p_i|k, \lambda_i) = \lambda_i$ and we use estimates for the probabilities from the 1993 to 1997 NCVS data as the $\lambda_i$'s. Using the model described in Section 3 and using the NCVS data from 1993 to 1997, we obtain estimates for $\tau, \rho, \delta$ and $\omega$ easily [see, e.g., Stasny and Coker (1997)] by

TABLE 8
*Comparison of the effects from spouse presence on rape reports for the years 1993–1997 and 1998–2004*

|     | 1993–1997 | | | 1998–2004 | |
|     | **Spouse present** | **Not present** | | **Spouse present** | **Not present** |
|-----|---|---|---|---|---|
| Yes | 9 | 211 | Yes | 9 | 152 |
| No  | 22502 | 76513 | No | 38376 | 120472 |

14 Q. YU, E. A. STASNY AND B. LIimplementing the EM algorithm. The parameter $k$ can be thought of as the prior sample size and it specifies the extent to which the estimator depends on the prior information. Here we propose an empirical-Bayesian estimate of $k$. A detailed explanation and assessment of this method can be found in Bishop, Fienberg and Holland (1975) and Carlin et al. (2000). Denote the Bayes estimate for $P$ to be $\hat{Q}$, which is the $E(P|k, \Lambda, X)$. Then the risk function is

$$R(\hat{Q}, P) = \left(\frac{n}{n+k}\right)^2 (1 - \|P\|^2) + \left(\frac{k}{n+k}\right)^2 n\|P - \Lambda\|^2,$$

where $\|P\|^2 = p_1^2 + p_2^2 + \cdots + p_t^2$. Differentiating the risk function with respect to $k$ and setting the resulting equation equal to 0 yields the estimate of $k$, $\hat{k} = (1 - \|P\|^2)/\|P - \Lambda\|^2$, that minimizes the risk $R(\hat{Q}, P)$. The optimal value of $k$ depends on the unknown value of $P$. If we use the MLE $\hat{P} = X/n$ to replace $P$, then the estimated optimal value of $k$ is

$$(4.1) \qquad \hat{k} = \left(n^2 - \sum_{i=1}^{t} x_i^2\right) \bigg/ \left(\sum_{i=1}^{t} x_i^2 - 2n \sum_{i=1}^{t} x_i \lambda_i + n^2 \sum_{i=1}^{t} \lambda_i^2\right).$$

The empirical-Bayesian estimate of $P$ is then

$$(4.2) \qquad P^* = n/(n+\hat{k})(X/n) + \hat{k}/(n+\hat{k})\Lambda.$$

From equation (4.2), we see that $\hat{k}$ determines how much the estimator depends on the prior information. In equation (4.1) we can rewrite the denominator as $\sum_{i=1}^{t}(x_i - n\lambda_i)^2$. Therefore, the more the prior information represents the current data (i.e., the closer $n\lambda_i$ is to the observed mean of $x_i$), the more the estimators depend on the prior distribution.

4.3. *An EMB algorithm.* The EMB algorithm is a variant of the EM algorithm [see, e.g., Dempster, Laird and Rubin (1977)]. In the EM algorithm, the M-step involves maximizing the complete data likelihood function to obtain the MLE for the parameters. In EMB algorithm, we add a B-step in which the $\hat{k}$ in equation (4.1) is calculated based on the last M-step, and then empirical-Bayesian estimates are obtained by minimizing the risk function. Thus, at the E-step, we fill in the unobserved data with estimates based on the values from the last B-step (those parameters using Bayesian estimates) or from the last M-step (for those using classical estimates). We illustrate the use of the EMB algorithm in the NCVS example. In the example the empirical-Bayesian estimates are obtained for $\tau, \rho$ and $\delta$.

Using the cell probabilities shown in Table 5 and the complete data from Table 4, subject to the constraint that $\sum_i \sum_j \omega_{ij} = 1$, the likelihood function of the complete data has a simple multiplicative form and can be split into five factors, each a function of only the $\pi, \tau, \rho, \delta$ and $\omega$ parameters. The likelihood function, written so that the functions of the five types of parameters are obvious, is proportional to the following function (a "+" in a subscript indicates summation over the corresponding index):

$$
\begin{aligned}
&\pi^{y_{2+++}}(1-\pi)^{y_{1+++}} \\
&\quad \times \rho^{y_{1111}+y_{2111}+y_{2131}}(1-\rho)^{y_{1121}+y_{2121}} \\
&\quad \times \delta^{y_{1211}+y_{2211}+y_{2231}}(1-\delta)^{y_{1221}+y_{2221}} \\
&\quad \times \tau^{y_{2111}+y_{2112}+y_{2211}+y_{2212}+y_{2311}+y_{2312}} \\
&\quad \times (1-\tau)^{y_{2131}+y_{2132}+y_{2231}+y_{2232}+y_{2331}+y_{2332}} \\
&\quad \times \omega_{11}^{y_{1111}+y_{1121}+y_{2111}+y_{2121}+y_{2131}} \omega_{12}^{y_{1211}+y_{1221}+y_{2211}+y_{2221}+y_{2231}} \\
&\quad \times \omega_{13}^{y_{1311}+y_{2311}+y_{2331}} \omega_{14}^{y_{1411}+y_{2411}} \omega_{15}^{y_{1511}+y_{2511}} \omega_{21}^{y_{1112}+y_{2112}+y_{2132}} \\
&\quad \times \omega_{22}^{y_{1212}+y_{2212}+y_{2232}} \omega_{23}^{y_{1312}+y_{2312}+y_{2332}} \omega_{24}^{y_{1412}+y_{2412}} \omega_{25}^{y_{1512}+y_{2512}} \\
&\equiv (1-\pi)^{y_{1+++}}\pi^{y_{2+++}}\rho^{a_1}(1-\rho)^{a_2} \\
&\quad \times \delta^{b_1}(1-\delta)^{b_2}\tau^{c_1}(1-\tau)^{c_2}\left\{\prod_{i=1}^{5}\prod_{j=1}^{2}\omega_{ij}^{y_{+i+j}}\right\}.
\end{aligned}
\tag{4.3}
$$

Because of the multiplicative form of this likelihood function, we can accomplish maximization separately for the $\tau, \rho, \delta$ and $\omega$ parameters. The closed form MLEs for these parameters are as follows:

$$
\begin{aligned}
\hat{\pi} &= y_{2+++}/(y_{1+++} + y_{2+++}) \\
&= (\text{\# phone interviews}) / (\text{\# phone + personal interviews}), \\
\hat{\rho} &= a_1/(a_1 + a_2), \\
\hat{\delta} &= b_1/(b_1 + b_2), \\
\hat{\tau} &= c_1/(c_1 + c_2), \\
\hat{\omega}_{ij} &= y_{+i+j}/y_{++++},
\end{aligned}
$$

where $a_1, a_2, b_1, b_2, c_1$ and $c_2$ are as defined in equation (4.3). Based on the MLEs, we determine the prior distribution for parameters by calculating the $k$'s through equation (4.1) that minimize the risk functions. We then obtain the empirical-Bayesian estimates $\tau^*, \rho^*$ and $\delta^*$ from equation (4.2). For example, to estimate $\rho$,

$$
\begin{aligned}
\hat{k}_\rho &= 2a_1a_2/(a_1^2 + a_2^2 - 2 \times (a_1 + a_2)(a_1\lambda_\rho + a_2(1-\lambda_\rho)) \\
&\quad + (a_1+a_2)^2(\lambda_\rho^2 + (1-\lambda_\rho)^2))
\end{aligned}
$$

and

$$\rho^* = (a_1 + a_2)/(a_1 + a_2 + \hat{k}_\rho) \times a_1/(a_1 + a_2) + \hat{k}_\rho/(a_1 + a_2 + \hat{k}_\rho)\lambda_\rho,$$

where $\lambda_\rho$ is the estimated value of $\rho$ using the 1993 to 1997 NCVS data. Note that the B-step does not add significant computational load to the EM algorithm, since each estimate has a closed form.

The E-step of the EMB algorithm is similar to that for the EM algorithm, except that we use different estimates of parameters to calculate the expectations of missing cells. In our example, the E-step consists of obtaining the expected cell counts for the complete data matrix (Table 5), given the observed data and the current estimates of the $\pi, \tau, \rho, \delta$ and $\omega$ parameters. These expectations are particularly simple in the case of discrete data [see Little and Rubin (2002)] and amount to proportionally allocating the $x_{ijk}$'s of the observed data as shown in Table 6 to the $y_{ijkl}$ cells of Table 4 according to the current parameter estimates. For example,

$$\hat{y}_{1121} = x_{141} \times \frac{(1-\pi^*)(1-\rho^*)\omega^*_{11}}{(1-\pi^*)(1-\rho^*)\omega^*_{11} + (1-\pi^*)(1-\delta^*)\omega^*_{12} + (1-\pi^*)\omega^*_{14}}.$$

Other expected cell counts can be found similarly and, hence, are not shown here.

The E-, M- and B-steps are repeated until parameter estimates have converged to the desired degree of accuracy, which in our case is when the sum of the relative differences of all estimated probabilities between two iterations is less than 0.0001. The code for implementing the EMB algorithm on the NCVS data to adjust for respondent's bias and estimate the rape and domestic violence rates in the years from 1998 to 2004 can be found in the supplemental file [Yu, Stasny and Li (2008)].

To estimate the variances of the estimators, we use the jackknife. More specifically, we look on each quarter between the years 1998 and 2004 as a sampling unit (SU). This results in 28 sampling units. Using all 28 SUs, we obtain the best estimate, say, $m$, for the parameter. By throwing out the first SU, we use the jackknife data set of 27 "resampled" SUs to get another estimate, say, $m_1$. In the next step a new reasmpling is performed with the second SU being deleted, and a new estimate $m_2$ is obtained. The process is repeated for each sample unit, resulting in a set of estimates, $m_i, i = 1, \ldots, 28$. The error for estimation is given by the formula

$$\sigma_j^2 = \frac{27 \sum_{i=1}^{28}(m_i - m)^2}{28}.$$

To calculate the $m_i$, we always use the EMB algorithm and the same prior information, since different observations in the future would not change the prior knowledge.

TABLE 9
*Observed data (frequencies) for crimes from NCVS 1993 to 1997*

|  | Personal interview | | Telephone Interview |
|---|---|---|---|
|  | **Spouse present** | **Spouse not present** |  |
| Rape | 9 | 211 | 314 |
| Domestic violence | 12 | 581 | 652 |
| Other assault | 81 | 861 | 1903 |
| Personal larceny | 16 | 117 | 323 |
| No crime | 22393 | 74954 | 303610 |

**EM estimates for crimes from 1993 to 1997**

| $\hat{\omega}_{ij}$ | **Spouse present** | **Spouse not present** |
|---|---|---|
| Rape | 0.000577 | 0.001964 |
| Domestic violence | 0.001374 | 0.004676 |
| Other assault | 0.002475 | 0.008421 |
| Personal larceny | 0.000255 | 0.000868 |
| No crime | 0.222448 | 0.756942 |
| $\hat{\pi}=0.76 \quad \hat{\rho}=0.14 \quad \hat{\delta}=0.07 \quad \hat{\tau}=0.53$ | | |

## 5. Results.

5.1. *Complexity of the computation.* To check the computational complexity and the efficiency of the estimates, we use both the frequentist method and the Bayesian method to obtain the estimates. Since each Bayesian estimate has a closed form, to get a Bayesian estimate does not add much computational load at each iteration. From the jackknife analysis, the EM algorithm in the frequentist method iterates 77.57 times on average to obtain converged estimates. For the Bayesian method, an average of 76.43 iterations in the EMB algorithm is required for convergence. We conclude that the Bayesian method does not add significant computational load as compared with the traditional method.

5.2. *Analysis with unweighted data.* We first fit the model described in Section 3 to the unweighted data from 1993 to 1997 and form the prior distributions. The original data from 1993 to 1997 and the estimates are shown in Table 9.

We then fit the empirical-Bayesian model to the data from Table 2. The estimates obtained from the procedure are shown in Table 10. Notice that the estimated probability of rape adjusting for the dampening effect of the presence of a spouse and using a telephone interview is $0.000326+0.001024 = 0.001350$. Thus, we estimate about 1.35 rapes per 1,000 women. This compares with a rate of 0.79 per 1,000 based on the raw data (Table 1). Simi-larly, the estimated probability of domestic violence is $0.000702 + 0.002205 = 0.002907$. This results in an estimate of 2.91 incidences of domestic violence per 1,000 women, which we compare to a rate of 1.66 per 1,000 based on the raw data (Table 10).

We estimate the probability that a crime (except for personal larceny) is not reported because the interview is conducted over the telephone is $1 - \tau^* = 0.37$. Thus, for interviews conducted over the telephone with women who are victims of any type of personal crime (except for personal larceny), we estimate that approximately 37% of the women did not report the victimization. For this estimate, $k/(n+k) = 21/(21+3708) = 0.0058$. That is, to get the estimate, we depend 0.58% on the prior information. Similarly, the probability that a rape is not reported because a spouse is present is about $1 - \rho^* = 0.86$, with $k/(n+k) = 1418/(1418+193) = 0.88$. That is, for interviews with women who are victims of rape and whose spouse was present during the interview, we estimate that 86% of the women did not report the victimization. To get this estimate, we depend 88% on the prior information. The probability that domestic violence is not reported because a spouse is present is about $1 - \delta^* = 0.92$, and $k/(k+n) = 227/(227+404)$ is 0.36. That is, for the women who are victims of domestic violence and whose spouse was present during the interview, we estimate that 92% of the women did not report the victimization and the estimate depends 36% on the prior information.

Table 11 shows the estimated parameters from the EM algorithm without using the prior information. Comparing the results in Table 10 and Table 11, we see the gains from the Bayesian model: the variances (as shown in parentheses) of the estimates from the Bayesian model are significantly lower than those from the classical method—$\mathrm{Var}(\hat{\rho})/\mathrm{Var}(\rho^*) = 6.625$ and $\mathrm{Var}(\hat{\delta})/\mathrm{Var}(\delta^*) = 1.86$. Bayesian and frequentist methods estimate about the same effect of a telephone interview on reporting crime, and have similar variances. This might be due to the fact that the Bayesian estimate of $\tau$ depends little on prior information. On average, the estimation of $\tau$ depends

TABLE 10
*EMB estimates for crimes from 1998 to 2004*

| $\hat{\omega}_{ij}$ | Spouse present | Spouse not present |
|---|---|---|
| Rape | 0.000326 | 0.001024 |
| Domestic violence | 0.000702 | 0.002205 |
| Other assault | 0.001363 | 0.004281 |
| Personal larceny | 0.000144 | 0.000453 |
| No crime | 0.238954 | 0.750549 |
| $\pi^* = 0.72$    $\rho^* = 0.14(0.0008)$    $\delta^* = 0.08(0.0007)$    $\tau^* = 0.63(0.0008)$ | | |

NOTE. Variances for the corresponding estimates are shown in parentheses.

TABLE 11
*EM estimates for crimes from 1998 to 2004 (not using the prior information)*

| $\hat{\omega}_{ij}$ | Spouse present | Spouse not present |
|---|---|---|
| Rape | 0.000367 | 0.001211 |
| Domestic violence | 0.000766 | 0.002528 |
| Other assault | 0.001474 | 0.004863 |
| Personal larceny | 0.000135 | 0.000446 |
| No crime | 0.229860 | 0.758346 |
| $\hat{\pi}=0.72$  $\hat{\rho}=0.16(0.0053)$  $\hat{\delta}=0.09(0.0013)$  $\hat{\tau}=0.61(0.0008)$ | | |

0.60% on prior information, while the estimation of $\rho$ depends 87.47% and the estimation of $\delta$ depends 38.87% on prior information.

In all, we conclude that to estimate respondent bias, the Bayesian model using prior information leads to more efficient estimates.

5.3. *Analysis with weight-adjusted data.* We then apply our model to the weight-adjusted NCVS data to determine if the use of sample-based weights leads to conclusions different from those based on the raw, unweighted data. We take the usual approach of standardizing the weights so that they sum to the actual sample size. Thus, a woman's standardized weight is her original sample weight divided by the total of the sample-based weights for all women in the analysis. (The weights used here are the cross-sectional weights developed to make the sample representative of the population of interest at the time of the survey. Because there are no longitudinal weights available for the NCVS, we use the cross-sectional weights to reflect the socio-economic and demographic makeup of the population while recognizing their limitations.) We use the weighted data from 1993 to 1997 to form the prior distributions. These weight-adjusted data and the estimates are summarized in Table 12.

We again estimate $\tau, \rho$ and $\delta$ using the EMB-algorithm described in Section 4. The parameter estimates for crimes are shown in Table 13. Table 14 compares the original crime rates and our estimated crime rates for both weighted and unweighted data. We note that the crime rates are generally estimated at a higher level when using the weighted data.

**6. Conclusions and future research.** We have shown that estimated rates of rape and domestic violence among women are increased under a model that considers gag factor effects in reporting such crimes based on the type of interview and who is present during the interview. Also, we used prior information to obtain more efficient estimates. We noticed that the type of interview and who is present during the interview may have different influences on different women. As reported by Stasny and Coker (1997), compared with women not reporting rape and domestic violence, those reporting

TABLE 12
*Weight-adjusted data (frequencies) for crimes from NCVS 1993 to 1997*

|  | Personal interview | | Telephone interview |
|---|---|---|---|
|  | **Spouse present** | **Spouse not present** |  |
| Rape | 8.96 | 224.36 | 337.35 |
| Domestic violence | 11.26 | 608.61 | 695.37 |
| Other assault | 84.99 | 937.59 | 1991.93 |
| Personal larceny | 15.10 | 121.16 | 323.33 |
| No crime | 22174.79 | 77079.18 | 301423.02 |

**EM estimates for crimes from 1993 to 1997**

| $\hat{\omega}_{ij}$ | **Spouse present** | **Spouse not present** |
|---|---|---|
| Rape | 0.000587 | 0.002079 |
| Domestic violence | 0.001383 | 0.004892 |
| Other assault | 0.002515 | 0.008895 |
| Personal larceny | 0.000249 | 0.000882 |
| No crime | 0.215688 | 0.762828 |
| $\hat{\pi} = 0.75$  $\hat{\rho} = 0.14$  $\hat{\delta} = 0.06$  $\hat{\tau} = 0.53$ | | |

were younger, had annual incomes of less than $15,000, were unemployed, rented rather than own their homes, were not currently married, and had moved more than five times in the last three years. An important area for future research is to account for some of these factors in the model. The problem with such analysis, of course, is that as other variables are used in creating cross-classified tables the data becomes very sparse, particularly in the cells involving reporting rape or domestic violence.

Because rape and domestic violence are relatively rare events, we have to combine information from a number of years. Thus, we do not obtain enough repeated measures for a panel analysis. The methods described in

TABLE 13
*EMB estimates for crimes from 1998 to 2004 (weight-adjusted data)*

**Estimates for crimes**

| $\hat{\omega}_{ij}$ | **Spouse present** | **Spouse not present** |
|---|---|---|
| Rape | 0.000372 | 0.001226 |
| Domestic violence | 0.000774 | 0.002555 |
| Other assault | 0.001481 | 0.004887 |
| Personal larceny | 0.000135 | 0.000446 |
| No crime | 0.229839 | 0.758285 |
| $\hat{\pi} = 0.72$  $\rho^* = 0.14(0.0002)$  $\delta^* = 0.07(0.0005)$  $\tau^* = 0.61(0.0025)$ | | |

TABLE 14
*Crime rates (# of crimes per 1000 people) for years 1998–2004*

|  | Original rates | Fitted rates | |
| --- | --- | --- | --- |
|  |  | Unweighted data | Weighted data |
| Rape | 0.79 | 1.35 | 1.60 |
| Domestic violence | 1.66 | 2.91 | 3.33 |
| Other assault | 4.14 | 5.64 | 6.37 |
| Personal larceny | 0.60 | 0.60 | 0.58 |
| No crime | 992.82 | 989.50 | 988.12 |

this paper are mainly used on contingency table analysis. If sufficient data are collected, we would like to implement our model in a panel analysis to discover how the crime rates change over time.

In our current analysis, we assume that if there is panel attrition in survey, the pattern of attrition is random. Brame and Piquero (2003) found that the pattern of panel attrition might be related to the interviewee's characteristics. In our future research, we want to explore the panel attrition. If it is nonrandom panel attrition, we want to adjust our analysis accordingly so that the estimation of crime rates would be more accurate.

Our model was developed to adjust for response biases caused by the mode of interviewing and the sensitive nature of questions in reporting rape and domestic violence. The models described in our paper may be useful in other survey sampling settings where some known factors may result in response bias. For example, one can imagine that reporting various sources of income could be biased because of who is present during the interview. Moreover, if data are collected repeatedly, the Bayesian method would be efficient because it incorporates the previously collected information into an important estimate. The EMB algorithm helps to make the computation easy. To implement our method, first identify which part of the information would not change over time and then build that part of the information into the prior distributions. Following the EMB algorithm, we could easily obtain the estimates of parameters that are of interest. The methods proposed here, therefore, can be easily applied in such cases.

**Acknowledgments.** The authors thank Donald Mercante, the editor, associate editor and referees for constructive comments and suggestions that helped to improve the presentation of the paper.

## SUPPLEMENTARY MATERIAL

**R-code of EMB algorithm to adjust for response bias in NCVS data for estimating rape and domestic violence rates** (doi: 10.1214/08-AOAS160SUPP; .txt).

Q. Yu  
School of Public Health  
Lousiana State University  
1615 Poydras Street  
Suite 1400  
New Orleans, Louisiana 70112  
USA  
E-mail: qyu@lsuhsc.edu  
URL: http://publichealth.lsuhsc.edu/Faculty_pages/qyu/index.html

E. A. Stasny  
Department of Statistics  
Ohio State University  
404 Cockins Hall  
1958 Neil Avenue  
Columbus, Ohio 43210  
USA  
E-mail: eas@stat.osu.edu

B. Li
Department of Experimental Statistics
Louisiana State University
Baton Rouge, Louisiana 70803
USA
E-mail: bli@lsu.edu
URL: http://www.stat.lsu.edu/faculty/li/